\documentclass[a4paper,11pt]{article}

\pdfoutput=1 

\usepackage{bbm}
\usepackage{jheppub} 
\usepackage[T1]{fontenc}
\usepackage{simplewick} 
\usepackage{makecell}	
\usepackage{tabu}		
\usepackage[makeroom]{cancel} 
\usepackage[normalem]{ulem} 
\usepackage[utf8]{inputenc}
\usepackage{amsmath,amsfonts,amsthm,amssymb} 
\usepackage{subcaption} 

\newcommand{\p}{\partial}
\newcommand{\eq}{&\quad}

\newcommand{\rig}{\right.}
\newcommand{\lef}{\left.}

\newcommand{\para}{\parallel}

\newcommand{\disc}{\text{disc}}

\newcommand{\mco}{{\mathcal{O}}}

\newcommand{\cG}{\mathcal{G}}

\newcommand{\R}{\mathbb{R}}
\newcommand{\Z}{\mathbb{Z}}
\newcommand{\1}{\mathbbm{1}}

\newcommand{\hD}{{\hat{\Delta}}}

\newcommand{\hga}{{\hat{\gamma}}}

\newcommand{\hO}{{\hat{\mathcal{O}}}}
\newcommand{\hp}{{\hat{\phi}}}

\newcommand{\f}{{(f)}}
\newcommand{\Df}{\Delta^{(f)}}

\newcommand{\al}{\alpha}

\newcommand{\de}{\delta}
\newcommand{\e}{\epsilon}
\newcommand{\ph}{\phi}
\newcommand{\g}{\gamma}

\newcommand{\la}{\lambda}
\newcommand{\m}{\mu}

\newcommand{\x}{\xi}

\newcommand{\De}{\Delta}

\newcommand{\G}{\Gamma}

\newcommand{\X}{\Xi}

\preprint{UUITP-04/23}

\title{\boldmath The discontinuity method in a BCFT}

\author{Alexander Söderberg Rousu}

\affiliation{Department of Physics and Astronomy,
	Uppsala University,\\
	Box 516,
	SE-751 20 Uppsala,
	Sweden}

\emailAdd{alexander.soderberg.rousu@gmail.com}

\makeatletter
\gdef\@fpheader{}
\makeatother

\abstract{
	We consider a conformal field theory in the presence of a boundary, and explain how two-point correlators of mixed bulk-local operators can be bootstrapped by exploiting the analytical structure of the conformal blocks. This yields the operator product expansion coefficients in either bootstrap channel. We apply this bootstrap technique to an $O(N)$-model in $d = 4 - \epsilon$ dimensions to find the $\phi - \phi^5$ bulk correlator and the corresponding operator product expansion coefficients upto $\mathcal{O}(\epsilon)$.

	Parts of this paper was first presented in my thesis \cite{SoderbergRousu:2023ucv}.
}

\begin{document} 
	
\newtheorem{defin}{Definition}
\newtheorem{thm}{Theorem}
\newtheorem{cor}{Corollary}
\newtheorem{pf}{Proof}
\newtheorem{nt}{Note}
\newtheorem{ex}{Example}
\newtheorem{ans}{Ansatz}
\newtheorem{que}{Question}
\newtheorem{ax}{Axiom}

\maketitle

\section{Introduction}


In this paper we consider a \textit{conformal field theory} (CFT) in the presence of a flat boundary: a codimension one defect where there is only a physical region on one side of it. This gives rise to a \textit{boundary CFT} (BCFT) preserving a codimension one conformal symmetry group: $SO(d, 1)$ (in Euclidean space). This means that boundary-local operators behave in a similar manner as in a \textit{homogeneous CFT} (without a defect) of codimension one. In particular, boundary one-point functions are trivial, and the boundary two-point functions are completely fixed (upto a field normalization constant).\footnote{The normalization of boundary-local fields is fixed by that for bulk-local fields (and vice versa).}

In a BCFT, there is the \textit{boundary operator product expansion} (BOE): the OPE between a bulk-local field and the boundary itself. For a scalar, it is on the form
\begin{equation} \label{BOE}
\begin{aligned}
\mco(x) &= \sum_{\hO}\frac{\mu^{\mco}{}_{\hO}}{|x_\perp|^{\De - \hD}}\hat{C}(x_\perp^2\p_\para^2)\hO(x_\para) \ , \quad \hat{C}(x) = \sum_{m\geq 0}\frac{x^m}{(-4)^mm! \left( \hD - \frac{d - 3}{2} \right)_m} \ ,
\end{aligned}
\end{equation}
where the exchanged boundary-local fields, $\hO$, are scalar primaries (annihilated by the generators of the special conformal transformations along the boundary) of scaling dimension $\hD$, and the differential operator $\hat{C}$ generates the towers of descendants. Note that the exchanged fields does not have any $SO(d - 1)$-spin. Such spinning operators only appear in the BOE of a bulk-local field with non-trivial $SO(d)$-spin, say $l$, wherein such case the BOE contains boundary-local operators with $SO(d - 1)$-spin $\hat{l} \leq l$ \cite{Billo:2016cpy, Lauria:2018klo}. 

Due to the BOE \eqref{BOE} (or the broken bulk conformal symmetry), the one-point functions in the bulk are no longer trivial, and are allowed to be non-zero for scalar fields \cite{Billo:2016cpy}. This has important physical consequences as it can break the global symmetries. E.g. the extraordinary phase transition for the $O(N)$-model near four dimensions where one component of the bulk scalar has a non-trivial one-point function \cite{domb2000phase, PhysRevB.47.5841, Shpot:2019iwk}.

Bulk-boundary two-point functions are fixed upto a BOE coefficient, which can be compared to the OPE coefficients that appear in the three-point functions in a homogeneous CFT. This means that the BOE coefficients is a new addition to the CFT data.

The two-point function, $\langle \mco_1(x)\mco_2(y)\rangle$, for two (different) bulk-local scalars is not fixed by conformal symmetry, and it is often difficult to find using standard Feynman diagram techniques (as this requires defect-defect, bulk-defect and bulk-bulk propagators). We will write $\langle \mco_1(x)\mco_2(y)\rangle$ in terms of a function $f(\x)$ which depend on the single cross-ratios $\x$
\begin{equation}  \label{bulk-bulk corr}
\begin{aligned}
\langle \mco_1(x)\mco_2(y)\rangle &= A_d\frac{|2\,x_\perp|^{\frac{\De_{21}^-}{2}}|2\,y_\perp|^{\frac{\De_{12}^-}{2}}}{|s|^{\De_{12}^+}}f(\x) \ , \quad \x = \frac{s^2}{4\,x_\perp y_\perp} \ .
\end{aligned}
\end{equation}
Here $A_d$ is a field normalization constant, $\De_{ij}^\pm \equiv \De_i \pm \De_j$ with $\De_i$ being the scaling dimension of $\mco_i$, $s^\m \equiv x^\m - y^\m$ is the distance between the bulk fields and $x_\perp$, $y_\perp$ are the coordinates orthogonal to the boundary. The cross-ratio $\x$ diverges as $x_\perp, y_\perp \rightarrow 0$.

Now we can approach the defect in two different ways using the OPE: either we first use the bulk OPE which allows us to express $\langle \mco_1(x)\mco_2(y)\rangle$ in terms of bulk one-point functions (which is fixed upto a constant), or we can apply the BOE to each bulk operator, which allows us to express $\langle \mco_1(x)\mco_2(y)\rangle$ in terms of the orthogonal defect-defect two-point functions. This yields a bootstrap equation \cite{Liendo:2012hy}
\begin{equation} \label{BCFT Bootstrap}
\begin{aligned}
f(\x) &= \sum_{\mco}\la^{\mco_1\mco_2}{}_\mco \m^\mco{}_\1 \cG_\text{bulk}(\De; \x) = \x^{\frac{\De_{12}^+}{2}}\sum_{\hO}\m^{\mco_1}{}_\hO\m^{\mco_2}{}_\hO\cG_\text{bndy}(\hD; \x) \ .
\end{aligned}
\end{equation}
Since we consider external scalars, the exchanged operators in both channels are scalars. This bootstrap equation looks similar to that for a four-point function in a homogeneous CFT \cite{BELAVIN1984333, Rattazzi:2008pe}, with the major difference that here there are two entirely different operators in the two channels: either bulk- or boundary-local operators. This means that the bootstrap equation for $\langle \mco_1(x)\mco_2(y)\rangle$ is not a result of crossing symmetry (of one OPE).

The conformal blocks, $\cG_\text{bulk}(\De; \x)$ and $\cG_\text{bndy}(\hD; \x)$, are known in closed form for any spacetime dimension $d$ \cite{McAvity:1995zd}
\begin{equation} \label{BCFT Blocks}
\begin{aligned}
\cG_\text{bulk}(\De; \x) &= \x^{\De/2}{}_2F_1\left( \frac{\De}{2}, \frac{\De}{2}, \De - \frac{d - 2}{2}, -\x \right) \ , \\
\cG_\text{bndy}(\hD; \x) &= \x^{-\hD}{}_2F_1\left( \hD, \hD - \frac{d - 2}{2}, 2\,\hD - d + 2, -\x^{-1} \right) \ .
\end{aligned}
\end{equation}
The bootstrap equation \eqref{BCFT Bootstrap} have been studied numerically for identical external scalars assuming only bulk-interactions in \cite{Gliozzi:2015qsa, Liendo:2012hy, Padayasi:2021sik} and only boundary-interactions in \cite{Behan:2020nsf, Behan:2021tcn}.


In a homogeneous CFT, it is possible to project out the OPE coefficients by studying the branch cuts of the bootstrap equation using the \textit{Lorentzian inversion formula} (LIF) \cite{Caron-Huot:2017vep, Simmons-Duffin:2017nub}. This assumes analyticity of the conformal blocks in the spin. In a similar manner, dispersion relations for the four-point functions have been found by exploiting the analytical structure of the bootstrap equation \cite{Bissi:2019kkx, Carmi:2019cub}. 

LIF's have also been developed for the bootstrap function of bulk two-point functions near a conformal defect of codimension strictly greater than one in \cite{Lemos:2017vnx, Liendo:2019jpu, Gimenez-Grau:2021wiv}. The main difference from the homogeneous case is that there is now one LIF for either bootstrap channel. In deriving these formulas, it is assumed that the bootstrap equation is analytic in either the spins of the bulk- or defect-local fields. Dispersion relations for the bulk two-point function near a defect of codimension strictly greater than one was found in \cite{Barrat:2022psm}, and for a boundary in \cite{Bianchi:2022ppi}.

For external scalars, the exchanged operators in the BCFT bootstrap equation are all scalars \cite{Liendo:2012hy}. The lack of spinning fields have important consequences for analytical bootstrap: since LIF's require analyticity in the spin, we cannot find one in a BCFT. Thus it is important to develop other analytical methods to bootstrap these theories. One such method is the functional bootstrap in \cite{Kaviraj:2018tfd, Mazac:2018biw}. 

Another method is the \textit{discontinuity method} developed in \cite{Bissi:2018mcq} for a boundary and extended in \cite{Dey:2020jlc} to work for interfaces (where there is a bulk theory defined on each side of it). This method has also been applied to a supersymmetric BCFT in \cite{Gimenez-Grau:2020jvf}, and it projects out the OPE coefficients from the bootstrap equation for two external scalars. In these works, the discontinuity method has only been applied to two-point functions with identical external scalars. In Sec. \ref{Sec: discontinuity} we will generalize it to mixed correlators where the external scalars might differ. We will also show that it is possible to project out both the bulk OPE, $\la^{\mco_1\mco_2}{}_\mco \m^\mco{}_\1$, and the BOE coefficients, $\m^{\mco_1}{}_\hO\m^{\mco_2}{}_\hO$, from the bootstrap eq. \eqref{BCFT Bootstrap}, depending on which branch cut we study of the conformal blocks. 

In Sec. \ref{Sec: discontinuity} we derive formulas for the OPE coefficients \eqref{OPE coeff} (our main result) in both channels, given in terms of the contribution of conformal blocks (\ref{BCFT Bootstrap 2}, \ref{Gb, Gi}) appearing at previous order in the expansion parameter. At the end of this Section we also show that it reproduces the correct BOE coefficients at $\mco(\e^2)$ in the $\ph - \ph$ correlator in $d = 4 - \e$. In Sec. \ref{Sec: mixed} we apply our formulas for the OPE coefficients to the $\ph - \ph^5$ correlator. We are able to resum the boundary-channel, giving us the correlator upto $\mco(\e)$. Our results on the CFT data is presented in Sec. \ref{Sec: CFT data}. We conclude in Sec. \ref{Sec: concl} with an outlook on future directions.



\section{The discontinuity method in a BCFT} \label{Sec: discontinuity}

In this Section we will explain the discontinuity method from \cite{Bissi:2018mcq}, and generalize it to work for mixed correlators in the bulk. We will also show how the BOE coefficients can be extracted from the bootstrap eq. \eqref{BCFT Bootstrap} in a similar way.

\begin{figure} 
	\centering
	\includegraphics[width=0.5\textwidth]{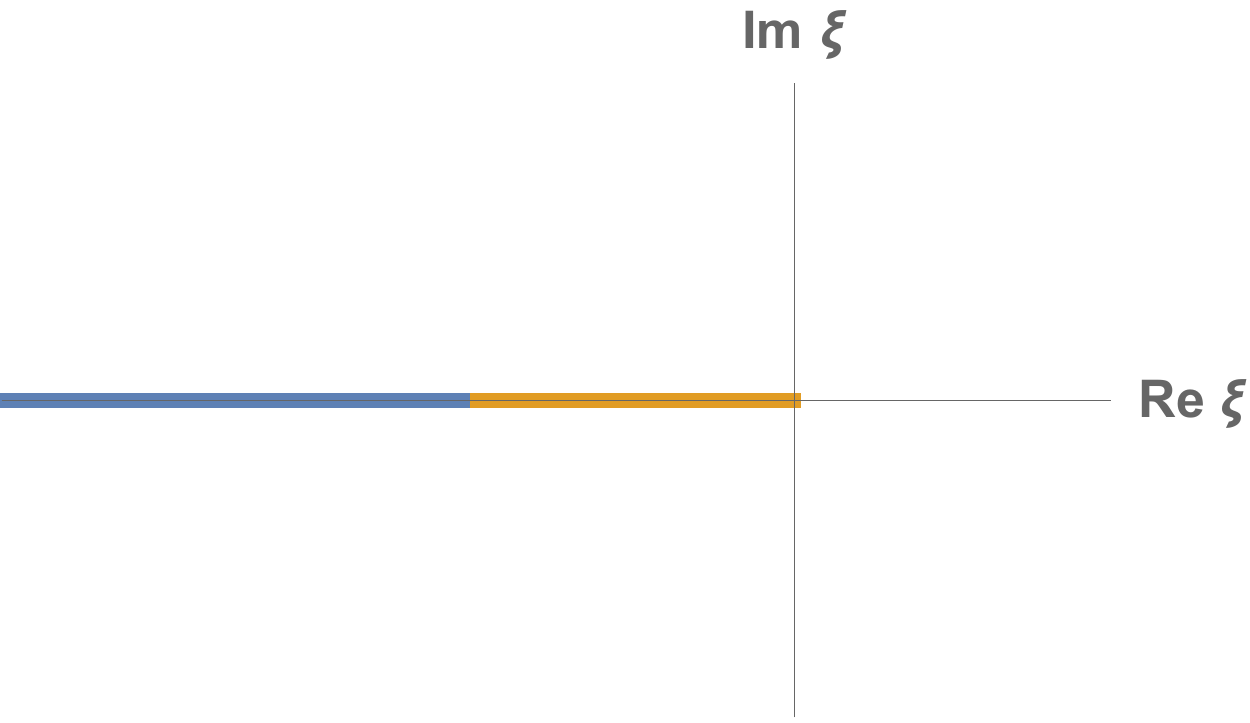}
	\caption{The branch cuts of the bootstrap eq. \eqref{BCFT Bootstrap}: the blue line is that of the bulk ${}_2F_1$, while the orange originates from the boundary ${}_2F_1$.}
	\label{Fig: bc}
\end{figure}

To start with, a hypergeometric function ${}_pF_q(..., \,z)$ has a branchcut along $z > 1$. This means that for the conformal blocks \eqref{BCFT Blocks}, the bulk-block, $\cG_\text{bulk}(\De; \x)$, has a branch cut along $\x < -1$ (this is the same as sending one of the external bulk operators to its \textit{mirror point} $x_\perp \rightarrow -x_\perp$) coming from the ${}_2F_1(..., -\x)$, and the boundary-block, $\cG_\text{bdy}(\hD; \x)$, has one along $-1 < \x < 0$ coming from the ${}_2F_1(..., -\x^{-1})$. In addition, the factors $\x^{\De/2}, \x^{-\hD}, \x^{\frac{\De_{12}^+}{2}}$ have branch cuts along the entire negative axis $\x < 0$. See Fig. \ref{Fig: bc} for an illustration on these branch cuts. In this Section we will show how the bulk OPE coefficients can be found from the branch cut along $\x < -1$, and the BOE coefficients from the branchcut along $-1 < \x < 0$.

We define the discontinuity of an arbitrary function $f(\x)$ along a real interval $\mathcal{I} \subset \R$ as
\begin{equation}
\begin{aligned}
\underset{{\xi \in \mathcal{I}}}\disc f(\x) = \lim\limits_{\al\rightarrow 0^+} \left[ f(\x + i\,\al) - f(\x - i\,\al) \right] \ , \quad \x \in \mathcal{I} \ .
\end{aligned}
\end{equation}
We note that the boundary-channel in the bootstrap eq. \eqref{BCFT Bootstrap} lacks a discontinuity along $\x < -1$ if $\frac{\De_{12}^+}{2} - \hD \in \Z$, or equivalently 
\begin{equation} \label{Bdy dim}
\begin{aligned}
\hD_m = \frac{\De_{12}^+}{2} + m \ , \quad m \in \Z \ .
\end{aligned}
\end{equation}
In the case of identical external scalars, $\De_1 = \De_2$, these operators correspond to the boundary-limit of normal derivatives acting on the scalar. These are the operators that appear in the conformal block decomposition in a \textit{generalized free field theory} (GFF).
The BOE coefficients are then given by eq. (B.45) in \cite{Liendo:2012hy}.

Likewise, we can move the $\x^{\frac{\De_{12}^+}{2}}$-factor in the bootstrap eq. \eqref{BCFT Bootstrap} to the bulk-channel, which then does not have a discontinuity along $-1 < \x < 0$ assuming the exchanged bulk operators have the scaling dimensions 
\begin{equation} \label{Bulk dim}
\begin{aligned}
\De_n = \De_{12}^+ + 2\,n \ , \quad n \in \Z \ .
\end{aligned}
\end{equation}
These operators correspond to scalar double traces: $\ph_1(\p_\m)^{2\,n}\ph_2$. In the case of $\De_1 = \De_2$, these are the operators appearing in the conformal block decomposition in a GFF, with the bulk OPE coefficients given by eq. (B.44) in \cite{Liendo:2012hy}.

In particular, if we assume exchanged operators of scaling dimensions (\ref{Bdy dim}, \ref{Bulk dim}), we can completely remove one side of the bootstrap eq. \eqref{BCFT Bootstrap} (in the free theory) by commuting the discontinuity (along either $\x < -1$ or $-1 < \x < 0$) with the infinite sums. Due to this wonderful analytical property, we will assume this spectrum of operators when we extract bulk OPE and BOE coefficients from the two bootstrap channels.

It is not fully known when we are allowed to commute the discontinuity with the infinite sums in the bootstrap eq. \eqref{BCFT Bootstrap}. In \cite{Bissi:2018mcq} (see its eq. (2.10)) this was discussed using the radial coordinates \cite{Lauria:2017wav} for boundary conformal blocks. In particular, the radial coordinate for the bulk-channel has a branch cut along $\x < -1$ while the one for the boundary has one along $-1 < \x < 0$. This means that it is important to double check the result against some known CFT data. E.g. in \cite{Bissi:2018mcq, Dey:2020jlc} the anomalous dimension of $\hp$ and $\p_\perp\hp$ was found (giving results consistent with the older literature) using the image symmetry of the conformal blocks
\begin{equation} \label{Im symm}
\begin{aligned}
\cG_{\text{bndy}}(\hD; e^{\pm\pi i}(\x + 1)) = e^{\mp\pi i\hD}\cG_{\text{bndy}}(\hD; \x) \ .
\end{aligned}
\end{equation}

\subsection{Discontinuity along $\x < -1$} \label{Sec: disc}

Firstly, let us study the discontinuity of the bulk-blocks \eqref{BCFT Blocks} along $\x < -1$ assuming exchanged scalar double traces \eqref{Bulk dim} 
\begin{align} \label{disc G bulk}
\underset{{\xi < -1}}\disc\cG_\text{bulk}(\De_n; \x) &= a_n\x^{\Df_\ph + 1 - \frac{\De_n}{2}}\times \nonumber \\
\eq\times{}_2F_1(\Df_\ph + 1 - \De_1 - n, \Df_\ph + 1 - \De_2 - n, \Df_\ph + 2 - \De_n, -\x) + \nonumber \\
\eq + b_n\x^{\frac{\De_n}{2}}{}_2F_1(\De_1 + n, \De_2 + n, \De_n - \Df_\ph, -\x) \ ,
\end{align}
where 
\begin{equation}
\begin{aligned}
a_n &= \frac{2\,\pi^2i\,\csc[\pi(\Df_\ph - \De_{12}^+)]\G_{\De_n - \Df_\ph}}{\G_{\De_1 + n}\G_{\De_2 + n}\G_{\De_1 + n - \Df_\ph}\G_{\De_2 + n - \Df_\ph}\G_{\Df_\ph + 2 - \De_n}} \ , \\
b_n &= e^{\pi\, i\, \De_{12}^+} - e^{2\,\pi\, i\,\Df_\ph} - 2i\frac{\sin[\pi(\De_1 + n - \Df_\ph)]\sin[\pi(\De_2 + n - \Df_\ph)]}{\sin[\pi(\De_{12}^+ - \Df_\ph)]} \ .
\end{aligned}
\end{equation}
Here $\G_n \equiv \Gamma(n)$ is a shorthand notation the Gamma function, and $\Df_\ph$ is a constant given by the scaling dimension of a fundamental field (without anomalous dimension)
\begin{equation} \label{Delta free}
\begin{aligned}
\Df_\ph = \frac{d - 2}{2} \ .
\end{aligned}
\end{equation}
We wish to find an orthogonality relation for \eqref{disc G bulk}, but this is difficult due to the two different ${}_2F_1$'s. However, the $b_n$-term vanishes in the case when we consider two identical external scalars with scaling dimension $\De_1 = \De_2 = \Df_\ph$. In such case it reduces to the discontinuity studied in eq. (4.11) of \cite{Bissi:2018mcq}. Alternatively, this term vanishes when $\Df_\ph$, $\De_1$ and $\De_2$ are all integers.

Let us consider one of these two scenarios, where we only need to focus on the $a_n$-term in \eqref{disc G bulk}. In such case we can use the following ${}_2F_1$-identity
\begin{equation} \label{2F1 id}
\begin{aligned}
{}_2F_1(a, b, c, z) = (1 - z)^{-a}{}_2F_1\left( a, c - b, c, \frac{z}{z - 1} \right) \ ,
\end{aligned}
\end{equation}
to rewrite the discontinuity as a Jacobi-polynomial
\begin{equation} \label{Bulk disc}
\begin{aligned}
\underset{{\xi < -1}}\disc\cG_\text{bulk}(\De_n; \x) &= c_n\frac{P_{\De_1 + n - \Df_\ph - 1}^{(\Df_\ph, -\De_{12}^-)}(t)}{(1 + t)^{\frac{\De_{12}^-}{2}}} \ ,
\end{aligned}
\end{equation}
with
\begin{equation}
\begin{aligned}
c_n = -\frac{\pi\,i^{2\,\Df_\ph - \De_n + 1}2^{\frac{\De_{12}^-}{2} + 1}\G_{\De_n - \Df_\ph}}{\G_{\De_1 + n}\G_{\De_2 + n - \Df_\ph}} \ , \quad t = -\frac{\x + 2}{\x} \ .
\end{aligned}
\end{equation}
Due to orthogonality of the Jacobi polynomial, the discontinuity \eqref{Bulk disc} of $\cG_\text{bulk}$ satisfies
\begin{equation} \label{Bulk orth}
\begin{aligned}
\int_{-1}^{+1}dt\,\X_m(t)\underset{{\xi < -1}}\disc\cG_\text{bulk}(\De_n; \x) = d_m\de_{mn} \ ,
\end{aligned}
\end{equation}
where 
\begin{equation}
\begin{aligned}
d_m &=  -\frac{\pi\,i^{2\,\Df_\ph - \De_m + 1}2^{\Df_\ph - \frac{\De_{12}^-}{2} + 2}\G_{\De_m - \Df_\ph + 1}}{\G_{\De_2 + m}\G_{\De_1 + m - \Df_\ph}} \ , \\ 
\X_m(t) &= \frac{(1 - t)^{\Df_\ph}}{(1 + t)^\frac{\De_{12}^-}{2}}P_{\De_1 + n - \Df_\ph - 1}^{(\Df_\ph, - \De_{12}^-)}(t) \ .
\end{aligned}
\end{equation}
This orthogonality relation is only valid if the coefficient, $d_m$, on the RHS of \eqref{Bulk orth} (which originates from the integration measure) is non-zero and convergent. E.g. in the case of two identical scalars with $\De_1 = \De_2 = \Df_\ph$, the discontinuity is orthogonal if $n \geq 1$. In which case the $\ph^2$-conformal block ($n = 0$) has a no branch cut along $\x < -1$.

\subsection{Discontinuity along $-1 < \x < 0$} \label{Sec: Bdy disc}

Let us now study the analytic structure of the boundary-channel, and consider the discontinuity of the boundary conformal block \eqref{BCFT Blocks} along $-1 < \x < 0$ assuming boundary operators of dimension \eqref{Bdy dim} is exchanged. In such case, we find again two terms with hypergeometric functions. One term vanishes if $\De_{12}^+$ and $\Df_\ph$ are both integers.\footnote{E.g. this occurs if $\De_1 = \De_2 = \De_\ph^\f$ in even dimensions.} Assuming this, we can again use the identity \eqref{2F1 id} to write the discontinuity as a Jacobi polynomial
\begin{equation} \label{Bdy orth}
\begin{aligned}
\underset{{-1 < \xi < 0}}\disc\cG_\text{bdy}(\hD_m; \x) &= \hat{c}_m\frac{P_{\hD_m - \Df_\ph - 1}^{(\Df_\ph, -\Df_\ph)}(u)}{(1 + u)^{\Df_\ph - \frac{\De_{12}^+}{2}}} \ ,
\end{aligned}
\end{equation}
with
\begin{equation}
\begin{aligned}
\hat{c}_m = -\frac{\pi\,i\,e^{\pi\,i(\Df_\ph - m)}2^{\Df_\ph - \frac{\De_{12}^+}{2} + 1}\G_{2(\hD_m - \Df_\ph)}}{\G_{\hD_m}\G_{\hD_m - 2\Df_\ph}} \ , \quad u = -2\,\x - 1 \ .
\end{aligned}
\end{equation}
The discontinuity \eqref{Bdy orth} satisfy the orthogonality relation
\begin{equation} \label{Bdy orth 2}
\begin{aligned}
\int_{-1}^{+1}du\,\hat{\X}_n(u)\underset{{-1 < \xi < 0}}\disc\cG_\text{bdy}(\hD_m; \x) = \hat{d}_n\de_{mn} \ ,
\end{aligned}
\end{equation}
where 
\begin{equation}
\begin{aligned}
\hat{d}_n &= \frac{\sqrt{\pi}\,i\,2^{\hD_n + n - \Df_\ph}e^{\pi\, i(\Df_\ph - n + 1)}\G_{\hD_n - \Df - \frac{1}{2}}}{\G_{\hD_n - \Df_\ph}} \ , \\
\hat{\X}_n(u) &= \frac{(1 - u)^{\Df_\ph}}{(1 + u)^\frac{\De_{12}^+}{2}}P_{\hD_m - \Df_\ph - 1}^{(\Df_\ph, - \Df_\ph)}(u) \ .
\end{aligned}
\end{equation}
Similar to the $\underset{{\xi < -1}}\disc\cG_\text{bulk}$, this relation is only valid if the coefficient, $\hat{d}_n$, is non-zero and convergent. In the case of $\De_1 = \De_2 = \Df_\ph$ in even dimensions, this is the case if $n \geq 2$. Then the conformal block for the $\hp$ ($m = 0$) and $\p_\perp\hp$ ($m = 1$) exchange have no branch cut along $-1 < \x < 0$.

\subsection{OPE coefficients}

We will now see how the orthogonality relations (\ref{Bulk orth}, \ref{Bdy orth}) can be used to project out the bulk OPE and BOE coefficients from the bootstrap eq. \eqref{BCFT Bootstrap}. The orthogonality relation only holds for free scaling dimensions (without taking into account the anomalous dimensions), and thus we first need to expand the CFT data around the free theory. To illustrate this we consider the $\e$-expansion, although in principle, the method work for other expansions as well, e.g. that around large $N$
\begin{equation} \label{exp 1}
\begin{aligned}
\De_1 &= \Df_1 + \e\,\g_1 + \mco(\e^2) \ , \\
\De_2 &= \Df_2 + \e\,\g_2 + \mco(\e^2) \ ,
\end{aligned}
\end{equation}
\begin{equation} \label{exp 2}
\begin{aligned}
\De_n &= \Df_n + \e\,\g_n + \mco(\e^2) \ , \\ 
\hD_m &= \hD_m^\f + \e\,\hga_m + \mco(\e^2) \ ,
\end{aligned}
\end{equation}
\begin{equation} \label{exp 3}
\begin{aligned}
\la^{\mco_1\mco_2}{}_{\mco_n} \m^{\mco_n}{}_\1 &= \la_n^\f + \e\,\de\la_n + \mco(\e^2) \ , \\
\m^{\mco_1}{}_{\hO_m}\m^{\mco_2}{}_{\hO_m} &= \m_m^{\f} + \e\,\de\m_m + \mco(\e^2) \ .
\end{aligned}
\end{equation}
We will then write the bootstrap eq. \eqref{BCFT Bootstrap} in the following way
\begin{equation} \label{BCFT Bootstrap 2}
\begin{aligned}
f(\x) &= G_b + H_b = G_i + H_i + \mco(\e^2) \ ,
\end{aligned}
\end{equation}
where $b$ stands for bulk and $i$ for interface (boundary). We let the anomalous dimensions be contained in $G_b$ and $G_i$ as well as the operators accompanying OPE coefficients of $\mco(\e)$ which does not have a branch cut along $\x < 0$ (such as $\hp$, $\p_\perp\hp$ and $\ph^2$ if $\De_1 = \De_2 = \Df_\ph$)
\begin{equation} \label{Gb, Gi}
\begin{aligned}
G_b &= \sum_{n}\la_n^\f \cG_\text{bulk}(\De_n; \x) + \e\sum_{n'}\de\la_{n'}\cG_\text{bulk}(\De_{n'}^\f; \x) \ , \\
G_i &= \sum_{m}\m_m^\f \cG_\text{bdy}(\hD_m; \x) + \e\sum_{m'}\de\m_{m'}\cG_\text{bulk}(\hD_{m'}^\f; \x) \ .
\end{aligned}
\end{equation}
On the other hand, we let $H_b$ and $H_i$ contain the operators at $\mco(\e)$ which have a non-trivial discontinuity along $\x < -1$ and $-1 < \x < 0$ respectively
\begin{equation}
\begin{aligned}
H_b &= \e\sum_{\tilde{n}}\de\la_{\tilde{n}}\cG_\text{bulk}(\De_{\tilde{n}}^\f; \x) \ , \\
H_i &= \e\sum_{\tilde{m}}\de\m_{\tilde{m}}\cG_\text{bulk}(\hD_{\tilde{m}}^\f; \x) \ .
\end{aligned}
\end{equation}
Note that the scaling dimensions (\ref{Bdy dim}, \ref{Bulk dim}) enter in $H_b$ and $H_i$ (upto $\mco(\e^0)$). If we now take the discontinuity of the bootstrap eq. \eqref{BCFT Bootstrap} along $\x < -1$ or $-1 < \x < 0$, and commute it with the infinite series in $H_b$ and $H_i$ we find
\begin{equation}
\begin{aligned}
\e\sum_{\tilde{n}}\de\la_{\tilde{n}}\underset{{\xi < -1}}\disc\cG_\text{bulk}(\De_n; \x) &= \underset{{\xi < -1}}\disc(G_i - G_b) \ , \\
\e\sum_{\tilde{m}}\de\m_{\tilde{m}}\underset{{-1 < \xi < 0}}\disc\cG_\text{bdy}(\hD_m; \x) &= \underset{{-1 < \xi < 0}}\disc(G_b - G_i) \ .
\end{aligned}
\end{equation}
By applying the orthogonality relations (\ref{Bulk orth}, \ref{Bdy orth}) we find the OPE coefficients at $\mco(\e)$
\begin{equation} \label{OPE coeff}
\begin{aligned}
\de\la_{\tilde{n}} &= \frac{d_{\tilde{n}}}{\e}\int_{-1}^{+1}dt\,\X_{\tilde{n}}(t)\underset{{\xi < -1}}\disc(G_i - G_b)\bigg|_{\x = -\frac{2}{t + 1}} \ , \\
\de\m_{\tilde{m}} &= \frac{\hat{d}_{\tilde{m}}}{\e}\int_{-1}^{+1}du\,\hat{\X}_{\tilde{m}}(u)\underset{{-1 < \xi < 0}}\disc(G_b - G_i)\bigg|_{\x = -\frac{u + 1}{2}} \ .
\end{aligned}
\end{equation}
If any of these formulas are used at $\mco(\e^k)$ with $k\geq 2$, then $\e \rightarrow \e^k$ in above formulas. That said, at some order in the expansion parameter we expect other operators than normal derivatives \eqref{Bdy dim} and scalar double traces \eqref{Bulk dim} (or to be precise: operators with the same scaling dimensions at $\mco(\e^0)$) to be exchanged. If so, above formulas are not valid anymore. 

It is worth mentioning that operators with scaling dimensions (\ref{Bdy dim}, \ref{Bulk dim}) will mix with other kinds of operators,\footnote{E.g. $\ph_1(\p_\m^{2})^n\ph_2$ might mix with $(\ph_1\ph_2)^{2\,n + 1}$ near four dimensions as they both have the same scaling dimension \eqref{Bulk dim} at $\mco(\e^0)$.} which means that above formulas actually give a sum of OPE coefficients. To go to higher orders, we have to solve the mixing problem by studying several different mixed bulk-bulk correlators. This is in general very difficult, and has only been done in very few theories. E.g. a $\mathcal{N} \geq 2$ supersymmetric theory in four dimensions \cite{Alday:2021ajh}.

The first of these formulas (for the bulk OPE coefficients) is the one used in \cite{Bissi:2018mcq, Dey:2020jlc, Gimenez-Grau:2020jvf} in the case of two identical external scalars $\De_1 = \De_2 = \Df_\ph$. However, the second formula for the BOE coefficients has not been used in the literature before.

Though the BOE coefficients were still found in \cite{Bissi:2018mcq, Dey:2020jlc} by resumming the bulk OPE coefficients,\footnote{Resumming conformal blocks is often difficult, and some methods to do so is in App. C of \cite{Dey:2020jlc}.} and then using the following orthogonality relation for the boundary-blocks themselves (and not their discontinuity)
\begin{equation*} 
\begin{aligned}
\de_{mn} &= \underset{|w| = \tilde{\e}}{\oint}\frac{dw}{2\,\pi\,i}w^{n - m - 1}{}_2F_1(1 - m, - m - \tfrac{d - 4}{2}, 2(1 - m), -w) {}_2F_1(n, n + \tfrac{d - 2}{2}, 2\,n, -w) \ .
\end{aligned}
\end{equation*}
Here we integrate over a small circle with infinitesimal radius $\tilde{\e} \ll 1$ s.t. we can apply the residue theorem to poles of the integrand at $w = \x^{-1} = 0$. This orthogonality relation was first found in \cite{Bissi:2018mcq}.

All and all, the discontinuity method provides us with the OPE coefficients in terms of the anomalous dimensions. This is a rewarding resolution to the bootstrap eq. \eqref{BCFT Bootstrap}, although it requires calculating difficult infinite sums in $G_b$ and $G_i$ if there are infinitely many operators at the previous orders in the expansion parameters (see Sec. 4.5 in \cite{Bissi:2018mcq}).


\subsubsection{BOE coefficients of the $\ph - \ph$ correlator}

In the remaining part of this Section, we will find the BOE coefficients at $\mco(\e^2)$ that appear in the $\ph - \ph$ correlator (when $\De_1 = \De_2 = \Df_\ph$) in  $d = 4 - \e$ using the second formula in \eqref{OPE coeff}. We will show that this results in an agreement with the results in \cite{Bissi:2018mcq}.

To start with, the bootstrap eq. \eqref{BCFT Bootstrap} is then solved by a finite number of operators in the free theory, and thus also at $\mco(\e)$ as the BOE coefficients enter squared in the bootstrap equation \cite{Liendo:2012hy}
\begin{equation} \label{phi phi Gb Gi}
\begin{aligned}
G_b &= 1 \pm (1 \pm \al\e + \e^2\de\la_{\ph^2})\cG_\text{bulk}(\De_{\ph^2}, \x) + \e\frac{\al}{2}\cG_\text{bulk}(\De_{\ph^4}; \x) \ , \\
G_i &= (1 \pm 1 + \e^2\de\mu_{\hp})\cG_\text{bdy}(\De_{\hp}; \x)  + \left( \frac{1 \mp 1}{2}\Df_{\ph} + \e\al + \e^2\de\mu_{\p_\perp\hp} \right) \cG_\text{bdy}(\De_{\p_\perp\hp}; \x) \ .
\end{aligned}
\end{equation}
Here $\De_0 \equiv \De_{\ph^2}$, $\De_1 \equiv \De_{\ph^4}$, $\hD_0 = \hD_\hp$, $\hD_1 \equiv \hD_{\p_\perp\hp}$ and $\al$ is a free parameter (that is related to the anomalous dimension of $\ph^2$). Neumann b.c. correspond to $+1$ and Dirichlet to $-1$. The expansion of these blocks are in App. A of \cite{Bissi:2018mcq}. The contributions from new OPE coefficients at $\mco(\e^2)$ are
\begin{equation}
\begin{aligned}
H_b &= \e^2\sum_{n \geq 1}\de\la_{n}\cG_\text{bulk}(2(n + 1); \x) \ , \quad H_i &= \e^2\sum_{m \geq 2}\de\m_{m}\cG_\text{bulk}(m + 1; \x) \ .
\end{aligned}
\end{equation}
By taking the discontinuity along $-1 < \x < 0$ of the difference $G_b - G_i$ we find the integrand of the BOE coefficients \eqref{OPE coeff}
\begin{equation} 
\begin{aligned}
\de\m_{m\geq 2} &= \frac{\sqrt{\pi}\,\G_m}{(-4)^m\G_{m - \frac{1}{2}}}\int_{-1}^{+1}du \left( a_1 \log\left(\frac{1 + u}{2}\right) + \rig \\
\eq\lef + a_2 \frac{1 - u}{1 + u} \log\left(\frac{1 - u}{2}\right) + a_3 + a_4 \frac{1 - u}{1 + u} \right) P_{m - 1}^{(+1, -1)}(u) \ ,
\end{aligned}
\end{equation}
with the coefficients
\begin{equation}
\begin{aligned}
a_1 &= \pm\al(2\,\al - 1) \ , \\
a_2 &= \al(2\,\al - \g_{\ph^4}^{(1)} + 1) \ , \\
a_3 &= \al(2(3 \pm 1)\al + 5 \mp 1) + \\
\eq - 4 (\al\,\g_{\ph^4}^{(1)} \pm 2\,\g_\ph^{(2)} \mp \g_{\ph^2}^{(2)} + (1\pm 1)\g_\hp^{(2)} - (1\mp 1)\g_{\p_\perp\hp}^{(2)} ) \ , \\
a_4 &= 2\,\g_\ph^{(2)} \ .
\end{aligned}
\end{equation}
Performing the integration over $t$ yields
\begin{align*}
\de\m_{m\geq 2} &= \frac{\sqrt{\pi}\,\G_m}{2^{2m - 1}(m - 1)\G_{m - \frac{1}{2}}} \bigg( \frac{(-1)^m(m^2 - m - 1)\al ( 2\al - \g_{\ph^4}^{(1)} + 1 ) }{m(m - 1)}  \mp \frac{\al(2\al - 1)}{m(m - 1)} + \nonumber \\
\eq + 2 \left( 1 \mp (-1)^m \right)\g_\ph^{(2)} + (-1)^m \left( \pm\g_{\ph^2}^{(2)} + (1 \pm 1)\hga_\hp^{(2)} - (1 \mp 1)\hga_{\p_\perp\hp}^{(2)} \right) \bigg) \ .
\end{align*}
By inserting the anomalous dimensions (see Sec. 4.2 in \cite{Bissi:2018mcq}), this result is in agreement with the corresponding BOE coefficients found in eq. (4.35) of \cite{Bissi:2018mcq}.

\section{The mixed $\ph - \ph^5$ correlator} \label{Sec: mixed}

For the rest of this paper we will apply the formulas \eqref{OPE coeff} to a mixed correlator in $d = 4 - \e$. We wish to avoid the issue of having an infinite amount of primaries at $\mco(\e^0)$. This occurs e.g. for the $\ph^2 - \ph^2$ correlator (see App. B.3 in \cite{Liendo:2012hy}). Thus we choose one of the external fields to be $\ph$. Since $\ph^3$ is a descendant of $\ph$ (as seen from the equation of motion if we assume a quartic interaction), the simplest non-trivial correlator we can consider with this operator is $\ph - \ph^5$.\footnote{Note that $\ph^5$ is the only primary of scaling dimension $\De = 5 + \mco(\e)$.}

\subsection{Free theory}

In the free theory, this correlator is given by
\begin{equation}
\begin{aligned}
\langle\ph^i(x)[(\ph^k)^2]^2\ph^j(y)\rangle = (N + 2)(N + 4)\de^{ij}\langle\ph^2(y)\rangle^2\langle\ph(x)\ph(y)\rangle \ .
\end{aligned}
\end{equation}
If we write this on the form \eqref{bulk-bulk corr}, with the constant
\begin{equation}
\begin{aligned}
A_d = \frac{(N + 2)(N + 4)}{(d - 2)^3S_d^3} \ , \quad S_d = \frac{2\,\G_{\frac{d}{2}}}{\pi^{\frac{d}{2}}} \ ,
\end{aligned}
\end{equation}
where $S_d$ is the solid angle in $d$-dimensions, then
\begin{equation} \label{phi - phi^5}
\begin{aligned}
f(\x) = \x^{2\,\De_\ph^\f} \left( 1 \pm \left( \frac{\x}{\x + 1} \right)^{\De_\ph^\f} \right) \ .
\end{aligned}
\end{equation}
Here $\pm$ denote Neumann/Dirichlet b.c.'s. By expanding in $\x$, we can decompose this function into the conformal blocks \eqref{BCFT Blocks}. Let us start with the bulk-channel. The $\x^{2\,\De_\ph}$-term is decomposed into
\begin{equation}
\begin{aligned}
\x^{2\,\De_\ph} = \la^{\ph\ph^5}{}_{\ph^4}\m^{\ph^4}{}_\1\cG_{\text{bulk}}(\De_{\ph^4}, \x) \ , \quad \la^{\ph\ph^5}{}_{\ph^4}\m^{\ph^4}{}_\1 = 1 \ ,
\end{aligned}
\end{equation}
with only $\ph^4$ being exchanged: $\De_{\ph^4} = 4\,\De_\ph$. On the other hand, the $\pm\x^{2\,\De_\ph}\left( \frac{\x}{\x + 1} \right)^{\De_\ph}$ is decomposed into an infinite amount of blocks, with $\De_n = 6\,\De_\ph + 2\,n$, for general $\De_\ph$
\begin{equation}
\begin{aligned}
\pm\x^{2\,\De_\ph}\left( \frac{\x}{\x + 1} \right)^{\De_\ph} &= \sum_{n\geq 0}\la^{\ph\ph^5}{}_{\mco_n}\m^{\mco_n}{}_\1\cG_{\text{bulk}}(\De_n; \x) \ , \\
\la^{\ph\ph^5}{}_{\mco_n}\m^{\mco_n}{}_\1 &= \pm\frac{(\De_\ph)_n(\De_\ph - \frac{d - 2}{2})_n(6\,\De_\ph -\frac{d}{2})_n}{(-4)^nn!(3\,\De_\ph - \frac{d}{4})_n(3\,\De_\ph - \frac{d - 2}{4})_n} \ .
\end{aligned}
\end{equation}
Note that there will be mixing among these operators in the quantized theory. If we specialize to $\De_\ph = \De_\ph^\f$ \eqref{Delta free} we find that only $\ph^6$ is exchanged: $\De_{\ph^6}^\f = 6\,\De_\ph^\f$
\begin{equation}
\begin{aligned}
\pm\x^{2\,\De_\ph^\f}\left( \frac{\x}{\x + 1} \right)^{\De_\ph^\f} &= \la^{\ph\ph^5}{}_{\ph^6}\m^{\ph^6}{}_\1\cG_{\text{bulk}}(\De_{\ph^6}; \x) \ , \quad \la^{\ph\ph^5}{}_{\ph^6}\m^{\ph^6}{}_\1 &= \pm 1 \ .
\end{aligned}
\end{equation}
All and all we find that only $\ph^4$ and $\ph^6$ is exchanged in the bulk-channel of \eqref{phi - phi^5}
\begin{equation}
\begin{aligned}
f(\x) = \cG_{\text{bulk}}(\De_{\ph^4}^\f; \x) \pm \cG_{\text{bulk}}(\De_{\ph^6}^\f; \x) \ .
\end{aligned}
\end{equation}
For the boundary-channel we can decompose the entire function $f(\x)$ (for general $\De_\ph$) right away in conformal blocks (by expanding in $\x$) to find an infinite tower of exchanged operators with $\hD_m = \De_\ph + m$
\begin{equation}
\begin{aligned}
f(\x) &= \x^{3\,\De_\ph}\sum_{m\geq 0}\m^{\ph}{}_{\hO_m}\m^{\ph^5}{}_{\hO_m}\cG_{\text{bdy}}(\hD_m; \x) \ , \\
\m^{\ph}{}_{\hO_m}\m^{\ph^5}{}_{\hO_m} &= \frac{(1 \pm 1) \G_{\De_\ph - \frac{d - 1}{2} + m} \left( \frac{\De_\ph}{2} \right)_m \left( \frac{\De_\ph + 1}{2} \right)_m \left( \De_\ph - \frac{d - 2}{2} \right)_m}{(2\,m)! \G_{\De_\ph - \frac{d - 1}{2} + 2\,m}} \ .
\end{aligned}
\end{equation}
If we specify to \eqref{Delta free} we find that either $\hp$ or $\p_\perp\hp$ is exchanged
\begin{equation}
\begin{aligned}
f(\x) &= \x^{3\,\De_\ph^\f} \left[ \m^{\ph}{}_{\hp}\m^{\ph^5}{}_{\hp} \cG_{\text{bdy}}(\De_\hp^\f; \x) + \m^{\ph}{}_{\p_\perp\hp}\m^{\ph^5}{}_{\p_\perp\hp} \cG_{\text{bdy}}(\De_{\p_\perp\hp}^\f; \x) \right] \ ,
\end{aligned}
\end{equation}
with the scaling dimensions
\begin{equation}
\begin{aligned}
\De_\hp^\f = \De_\ph^\f \ , \quad \De_{\p_\perp\hp}^\f = \De_{\ph}^\f + 1 \ ,
\end{aligned}
\end{equation}
and BOE coefficients
\begin{equation}
\begin{aligned}
\m^{\ph}{}_{\hp}\m^{\ph^5}{}_{\hp} &= 1 \pm 1 \ , \quad \m^{\ph}{}_{\p_\perp\hp}\m^{\ph^5}{}_{\p_\perp\hp} = \frac{1 \mp 1}{2}\De_\ph^\f \ . 
\end{aligned}
\end{equation}
These BOE coefficients are the same as in the $\ph - \ph$ correlator \eqref{phi phi Gb Gi}, meaning that in the free theory
\begin{equation}
\begin{aligned}
\m^{\ph^5}{}_{\hp} &= \m^{\ph}{}_{\hp} \ , \quad \m^{\ph^5}{}_{\p_\perp\hp} = \m^{\ph}{}_{\p_\perp\hp} \ . 
\end{aligned}
\end{equation}
To summarize: $\ph^4$ and $\ph^6$ are the two operators exchanged in the bulk-channel, while in the boundary-channel only one operator ($\hp$ or $\p_\perp\hp$ depending on the b.c.) is exchanged.

\subsection{Interacting theory}

In the interacting theory we expect an infinite amount of operators to appear in the two channels.\footnote{Note that the BOE coefficients enter squared in the $\ph - \ph$ correlator, which has an infinite tower of exchanged operator at $\mco(\e^2)$.} Let us expand the CFT data in $\e$ as in (\ref{exp 1}, \ref{exp 2}, \ref{exp 3}) with the exchanged scaling dimensions (\ref{Bdy dim}, \ref{Bulk dim}). In these notations only $n = -1$ ($\ph^4$) and $n = 0$ ($\ph^6$) appear in the bulk-channel at $\mco(\e^0)$, and either $m = -2$ ($\hp$) or $m = -1$ ($\p_\perp\hp$) is exchanged at $\mco(\e^0)$ in the boundary-channel. For the external $\ph$ and $\ph^5$ operators: only the bulk conformal blocks accompanied by $\de\la_n$ for $n \in \Z_{\geq 1}$ have a non-trivial discontinuity along $\x < -1$, and the boundary conformal blocks times $\de\m_m$ for $m \in \Z_{\geq 0}$ have a non-trivial discontinuity along $-1 < \x < 0$. 

This means that we will include the contribution from $\ph^{2}$, $\ph^{4}$ and $\ph^{6}$ ($n \in \{-2, -1, 0\}$) in $G_b$.
Note that $\ph^2$ is not exchanged in the free theory, but in general it might still appear at $\mco(\e)$. Let us assume that both $\hp$ and $\p_\perp\hp$ appear in $G_i$ ($m \in \{-2, -1\}$).\footnote{Later in we will see that only $\hp$/$\p_\perp\hp$ appear for Neumann/Dirichlet b.c.'s, which is what we expect from the $\ph - \ph$ correlator \cite{Bissi:2018mcq}.} This gives us
\begin{equation*}
\begin{aligned}
G_b &= \cG_\text{bulk}(6\,\De_\ph^\f - 2 + \e\,\g_{-1}; \x) + \cG_\text{bulk}(6\,\De_\ph^\f + \e\,\g_0; \x) + \e\sum_{n = -2}^0\de\la_{n}\cG_\text{bulk}(6 + 2\,n; \x) \ , \\
G_i &= \cG_\text{bdy}(3\,\De_\ph^\f - 2 + \e\,\hga_{-2}; \x) + \cG_\text{bdy}(3\,\De_\ph^\f - 1 + \e\,\hga_{-1}; \x) + \e\sum_{m = -2}^{-1}\de\m_{m}\cG_\text{bdy}(3 + m; \x) \ .
\end{aligned}
\end{equation*}
These can be expanded in $\e$ using the Mathematica package 'HypExp' \cite{Huber:2005yg, Huber:2007dx}, giving us the results in App. \ref{App: exp}.
$H_b$ and $H_i$ are on the other hand given by
\begin{equation} \label{Hb Hi}
\begin{aligned}
H_b &= \e\sum_{n \geq 1}\de\la_{n}\cG_\text{bulk}(6 + 2\,n; \x) \ , \quad H_i = \e\sum_{m \geq 0}\de\m_{m}\cG_\text{bdy}(3 + m; \x) \ .
\end{aligned}
\end{equation}
The formulas at \eqref{OPE coeff} gives us
\begin{equation} \label{Bulk OPE Coeff} 
\begin{aligned}
\de\la_{n\geq 1} &= \frac{(-1)^n\G_n\G_{n + 5}}{2\,\G_{2(n + 2)}}\int_{-1}^{+1}dt \left( a_1 + a_2(t + 1) + a_3(t + 1)^4 \right) P_{n - 1}^{(1, 4)}(t) \\ 
&= \frac{ \left( 6 - (-1)^n(n + 1)_3 \right) a_1 + 6 \left( 2 \left( 1 - (-1)^n \right) - n \right) a_2 + 96\,a_3}{3(n + 4)} \ ,
\end{aligned}
\end{equation}
\begin{equation} \label{BOE Coeff} 
\begin{aligned}
\de\m_{m\geq 0} &= -\frac{\sqrt{\pi}\,\G_{m + 2}}{(-4)^{m + 2}\G_{m - \frac{3}{2}}}\int_{-1}^{+1}du \left( b_1 + b_2 \frac{1 - u}{1 + u} \right) P_{m}^{(+1, -1)}(u) \\
&= m!\frac{(-1)^{m + 1}b_1 + b_2}{4^{m + 1}(\frac{3}{2})_m} \ ,
\end{aligned}
\end{equation}
with the coefficients
\begin{equation} \label{Coeff a}
\begin{aligned}
a_1 &= -(1 \pm 1)(\g_\ph - \hga_{-2} + 1) \ , \\
a_2 &= \frac{2(\g_\ph + 1) - (1 \pm 1)\hga_{-2} - (1 \mp 1)\hga_{-1}}{4} \ , \\
a_3 &= \pm\frac{\g_\ph - \g_{\ph^5} + \g_0}{32} \ ,
\end{aligned}
\end{equation}
\begin{equation} \label{Coeff b}
\begin{aligned}
b_1 &= \pm(\g_\ph + \g_{\ph^5} - \g_0 + 2) - (1 \pm 1)\hga_{-2} + (1 \mp 1)\hga_{-1} \ , \\
b_2 &= -\g_\ph - \g_{\ph^5} + \g_{-1} - 1 \ .
\end{aligned}
\end{equation}
We can resum the boundary-channel, $H_i$, in \eqref{Hb Hi} with the BOE coefficients \eqref{BOE Coeff} using the following integral representation of the ${}_2F_1$
\begin{equation} 
\begin{aligned}
{}_2F_1(a, b, c, z) = \frac{\G_c}{\G_b\G_{c - b}}\int_{0}^{1}dx\frac{x^{b - 1}(1 - x)^{c - b - 1}}{(1 - x\,z)^a} \ .
\end{aligned}
\end{equation}
If we first perform the summation over $m$ in $H_i$, and then integrate over $x$, we find
\begin{equation} 
\begin{aligned}
H_i = \frac{\e\,\x^2}{2} \left[ \left( \frac{\x}{\x + 1}a_1 + a_2 \right) \log\left( 1 + \frac{1}{\x} \right) - \frac{a_1 + a_2}{\x + 1} \right] \ .
\end{aligned}
\end{equation}
Note that this function does not have any branch cuts along $\x < -1$ as expected. Following step 1.e. in Sec. 2.1 of \cite{Dey:2020jlc} we see that the contribution (resummation) from even and odd blocks (w.r.t. $m$) in $H_i$ trivially satisfies image symmetry \eqref{Im symm}. Using the $H_i$ above, we can find the full correlator, $f(\x)$, and $H_b$ from the bootstrap eq. \eqref{BCFT Bootstrap 2}
\begin{equation} 
\begin{aligned}
f(\x) &= G_i + H_i \ , \quad H_b = f(\x) - G_b \ .
\end{aligned}
\end{equation}
With these we can make several checks. Firstly, $H_b$ does not have any branch cuts along $-1 < \x < 0$. Secondly, by imposing the b.c.'s on the $\ph - \ph^5$ correlator (using $f(\x)$) we find that for Neumann b.c.\footnote{Note that there is a pole in the boundary-limit which corresponds to the $\hp$-exchange. This term is not zero since we expect $\hp$ to be exchanged.}
\begin{equation} \label{BOE Coeff Neu} 
\begin{aligned}
\de\m_{-2} = 0 \ ,
\end{aligned}
\end{equation}
while for Dirichlet b.c.
\begin{equation} \label{BOE Coeff Dir}
\begin{aligned}
\de\m_{-1} = 0 \ .
\end{aligned}
\end{equation}
These are the expected results.

\subsection{CFT data} \label{Sec: CFT data}

To summarize, from conformal bootstrap of the $\ph - \ph^5$ correlator we find the OPE coefficiients to be given by (\ref{Bulk OPE Coeff}, \ref{BOE Coeff}) with no constraints on the anomalous dimensions. Moreover, we find no constraints on the OPE coefficients at $\mco(\e)$ for the bulk exchange of $\ph^2$ ($\de\la_{-2}$), $\ph^4$ ($\de\la_{-1}$) and $\ph^6$ ($\de\la_{0}$). I.e.
\begin{equation}
\begin{aligned}
\left(\la^{\ph\ph^5}{}_{\ph^2}\m^{\ph^2}{}_\1\right)^\pm &= \mco(\e) \ , \\ \left(\la^{\ph\ph^5}{}_{\ph^4}\m^{\ph^4}{}_\1\right)^\pm &= 1 + \mco(\e) \ , \\ \left(\la^{\ph\ph^5}{}_{\ph^6}\m^{\ph^6}{}_\1\right)^\pm &= \pm 1 + \mco(\e) \ ,
\end{aligned}
\end{equation}
where $+$ corresponds to Neumann b.c.'s, and $-$ to Dirichlet. For Neumann/Dirichlet b.c.'s we find that $\p_\perp\hp$/$\hp$ is not exchanged (\ref{BOE Coeff Neu}, \ref{BOE Coeff Dir}). For Neumann b.c.'s we can use as input from the $\ph - \ph$ correlator that there are no correction to the $\hp$ BOE coefficient \cite{Bissi:2018mcq} (which does not mix with any other boundary operators) to find
\begin{equation}
\begin{aligned}
\left(\m^{\ph}{}_{\hp}\m^{\ph^5}{}_{\hp}\right)^+ &= 2 + \mco(\e^2) \ .
\end{aligned}
\end{equation}
For Dirichlet b.c.'s we are not able to use any such input from the $\ph - \ph$ correlator, and thus we are not able to find the correction of the $\p_\perp\hp$ BOE coefficient at $\mco(\e)$
\begin{equation} \label{Dir BOE}
\begin{aligned}
\left(\m^{\ph}{}_{\p_\perp\hp}\m^{\ph^5}{}_{\p_\perp\hp}\right)^- = 1 + \mco(\e) \ . 
\end{aligned}
\end{equation}
We will now write out the non-trivial OPE coefficiients (\ref{Bulk OPE Coeff}, \ref{BOE Coeff}) at $\mco(\e)$ using as input the anomalous dimensions of the exchanged operators in the bulk \cite{Rychkov:2015naa}
\begin{equation} \label{Input 1}
\begin{aligned}
\g_{\ph} = 0 \ , \quad \g_{\ph^5} = \frac{2(N + 14)}{N + 8} \ , \quad \g_{-1} = 3 \ , \quad \g_0 = \frac{3(N + 14)}{N + 8} \ .
\end{aligned}
\end{equation}
We remind the reader that $\g_{-1}$ is the anomalous dimension of $\ph^4$ and $\g_{-2}$ that of $\ph^6$. On the boundary we use as input \cite{McAvity:1995zd}
\begin{equation} \label{Input 2}
\begin{aligned}
\hga_{-2} \ ,\  \hga_{-1} = \frac{N + 14}{2(N + 8)} \ .
\end{aligned}
\end{equation}
Here $\hga_{-2}$ is the anomalous dimension of $\hp$ and $\hga_{-1}$ that of $\p_\perp\hp$.\footnote{These anomalous dimensions correspond to $\De_{\ph^4} = 4 + \mco(\e^2)$, $\De_{\ph^6} = 6 + \frac{18\,\e}{N + 8} + \mco(\e^2)$, \\
	$\De_\hp = 1 - \frac{N + 5}{N + 8}\e + \mco(\e^2)$ and $\De_{\p_\perp\hp} = 2 - \frac{N + 5}{N + 8}\e + \mco(\e^2)$.}
In the bulk-channel there is mixing among operators of scaling dimensions
\begin{equation}
\begin{aligned}
\De_{n\geq 1} = 2(3 + n) + \mco(\e) \ .
\end{aligned}
\end{equation}
The corresponding sum of bulk OPE coefficients (\ref{Bulk OPE Coeff}, \ref{Coeff a}) are given by (using the anomalous dimensions (\ref{Input 1}, \ref{Input 2}) as input)
\begin{equation*}
\begin{aligned}
(\la^{\ph\ph^5}{}_{\mco_n}\m^{\mco_n}{}_\1)_{n\geq 1}^+ &= \e\frac{(n - 1)!(n + 3)!}{(2\,n + 3)!}\frac{72(-1)^n + (N + 2)(n + 2)(2\,n(n + 4) + 3)}{12(N + 8)} + \mco(\e^2) \ , \\
(\la^{\ph\ph^5}{}_{\mco_n}\m^{\mco_n}{}_\1)_{n\geq 1}^- &= -\e\frac{(n - 1)!(n + 3)!}{(2\,n + 3)!}\frac{24(-1)^n + (N + 2)(n + 2)}{4(N + 8)} + \mco(\e^2) \ ,
\end{aligned}
\end{equation*}
where $+$ corresponds to Neumann b.c.'s, and $-$ to Dirichlet b.c.'s. Note that $(\la^{\ph\ph^5}{}_{\mco_n}\m^{\mco_n}{}_\1)_{n\geq 1}^+$ is always positive.

In the boundary-channel there is mixing among operators of scaling dimensions
\begin{equation}
\begin{aligned}
\hD_{m\geq 0}^+ = 2\,m + 3 + \mco(\e) \ , \quad \hD_{m\geq 1}^- = 2\,m + 3 + \mco(\e) \ .
\end{aligned}
\end{equation}
Note that only even/odd boundary operators (w.r.t. $m$) appear for Neumann/Dirichlet b.c.'s. The corresponding sum of BOE coefficients (\ref{BOE Coeff}, \ref{Coeff b}) are given by (using the anomalous dimensions (\ref{Input 1}, \ref{Input 2}) as input)
\begin{equation}
\begin{aligned}
(\m^{\ph}{}_{\hO_m}\m^{\ph^5}{}_{\hO_m})_{m\geq 0}^+ &= \frac{3\sqrt{\pi}\,\e}{N + 8}\frac{(2\,m)!}{16^m\G_{2\,m + \frac{3}{2}}} + \mco(\e^2) \ , \\
(\m^{\ph}{}_{\hO_m}\m^{\ph^5}{}_{\hO_m})_{m\geq 1}^- &= \frac{12\sqrt{\pi}\,\e}{N + 8}\frac{(2\,m - 1)!}{16^m\G_{2\,m + \frac{1}{2}}} + \mco(\e^2) \ .
\end{aligned}
\end{equation}
Here $\G_x \equiv \G(x)$ is the Gamma-function. 
Both of these are positive which is what we expect due to unitarity.

The $\ph - \ph^5$ correlator (on the form \eqref{bulk-bulk corr}) for Neumann b.c.'s is given by
\begin{equation}
\begin{aligned}
f^+(\x) &= \x^2 \left[ 1 + \frac{\x}{\x + 1} + \frac{\e}{N + 8} \left( 9\frac{\x}{\x + 1}\log\x + \rig\rig \\
\eq\lef\lef + \left( \frac{N + 14}{2} + \frac{N - 4}{2}\frac{\x}{\x + 1} \right) \log(\x + 1) \right) \right] + \mco(\e^2) \ ,
\end{aligned}
\end{equation}
while for Dirichlet b.c.'s it is
\begin{equation}
\begin{aligned}
f^-(\x) &= \frac{\x^2}{\x + 1} \left[ (\m^{\ph}{}_{\p_\perp\hp}\m^{\ph^5}{}_{\p_\perp\hp})^- - \frac{\e}{N + 8} \left( \frac{N + 26}{2} + 9\,\x\log\x + \rig\rig \\
\eq\lef\lef + \left( \frac{N + 14}{2} + 9\,\x \right) \log(\x + 1) \right) \right] + \mco(\e^2) \ ,
\end{aligned}
\end{equation}
where the BOE coefficient \eqref{Dir BOE} for the $\p_\perp\hp$-exchange is not determined at $\mco(\e)$.

\section{Conclusion} \label{Sec: concl}

In this paper we derived the formulas \eqref{OPE coeff} for the bulk OPE and the BOE coefficients that appear in the bootstrap eq. \eqref{BCFT Bootstrap} of a mixed bulk correlator near a conformal boundary. This was done by exploiting the analytical properties of the conformal blocks \eqref{BCFT Blocks}. We also showed how the formula for the BOE coefficients reproduce the correct result from \cite{Bissi:2018mcq}. After this we applied our formulas to the mixed $\ph - \ph^5$ correlator, which resulted in the OPE coefficients and the correlator in Sec. \ref{Sec: CFT data}.

In particular, we found that for Neumann b.c.'s all of the OPE coefficients (both in the bulk- and the boundary-channel) in the $\ph - \ph^5$ correlator are positive. Under the strong assumption that this holds to all orders in $\e$, it would be interesting to study the corresponding bootstrap equation numerically. See e.g. \cite{Liendo:2012hy}.


Another interesting direction would be to apply the discontinuity method presented in this work to other kinds of correlators. E.g. the bulk two-point function with Noether currents or stress-energy tensors \cite{Herzog:2017xha}, correlators involving fermions \cite{Herzog:2022jlx} or bulk-bulk-boundary three-point functions \cite{Behan:2020nsf}.

An even more ambitious direction would be to investigate the possibilities of finding formulas for the anomalous dimensions from the analytic structure of the conformal blocks \eqref{BCFT Blocks}. This would be of outmost importance in the case when there is an infinite amount of operators already in the first order of the expansion parameter, e.g. the $\ph - \ph$ correlator in $d = 3$ (exactly) expanded around large $N$ (see Sec. B.7 in \cite{Liendo:2012hy}). In such case we have the expansions of the conformal blocks
\begin{equation} 
\begin{aligned}
\cG_\text{bulk}(\De; \x) &= \cG_\text{bulk}(\De^{(0)}; \x) + \frac{\g^{(1)}}{N}\p_{\De}\cG_\text{bulk}(\De^{(0)}; \x) + \mco(N^{-2}) \ , \\
\cG_\text{bdy}(\hD; \x) &= \cG_\text{bdy}(\hD^{(0)}; \x) + \frac{\hga^{(1)}}{N}\p_{\hD}\cG_\text{bdy}(\hD^{(0)}; \x) + \mco(N^{-2}) \ .
\end{aligned}
\end{equation}
If we were able to express $\p_{\De}\cG_\text{bulk}(\De^{(0)}; \x)$ and/or $\p_{\hD}\cG_\text{bdy}(\hD^{(0)}; \x)$ in terms of e.g. hypergeometric functions, we could in turn study their analytic structure to see if it is possible to project out the anomalous dimensions, $\g^{(1)}$ and/or $\hga^{(1)}$, from the bootstrap equation.

\newpage

\section*{Acknowledgement}

I would like to express my gratitude to Agnese Bissi for several enriching discussions on boundaries. I also thank everyone that went to my public defence of my thesis \cite{SoderbergRousu:2023ucv}, where most results in this paper was first presented. 
This project was funded by Knut and Alice Wallenberg Foundation grant KAW 2021.0170, VR grant 2018-04438 and Olle Engkvists Stiftelse grant 2180108.

\appendix

\section{Expansion of conformal blocks} \label{App: exp}

If we expand the functions $G_b$ and $G_i$ in $\e$ we find \cite{Huber:2005yg, Huber:2007dx}
\begin{equation}
\begin{aligned}
G_b &= \x^2 \left( 1 \pm \frac{\x}{\x + 1} \right) + \e \left( \x(3\,\x + 1)\de\la_{-2} + \x^2\de\la_{-1} + \frac{\x^3}{\x + 1}\de\la_{0} + \rig \\
\eq - \frac{\g_\ph - \g_{\ph^5} - \g_{-1} - 1}{6} \frac{\x^3}{\x + 1} \pm \frac{\g_\ph - \g_{\ph^5} + \g_0}{24} \frac{3\,\x^3 - 4\,\x^2 + 6\,\x - 12}{\x + 1} + \\
\eq + \left( (\g_{-1} - 3) \pm (\g_0 - 3)  \frac{\x^2}{\x + 1} \right)  \frac{\x\,\log\x}{2} + \\
\eq\lef - \left[ (\g_\ph - \g_{\ph^5} + \g_{-1} - 1) \mp \left( \x^3 + (\g_\ph - \g_{\ph^5} + \g_0) \frac{\x^4 + 1}{\x(\x + 1)} \right) \right] \frac{\log(\x + 1)}{2} \right) \ , 
\end{aligned}
\end{equation}
\begin{equation}
\begin{aligned}
G_i &= \x^2 \left( 1 \pm \frac{\x}{\x + 1} \right) + \e \left( \x^2 \left( 1 + \frac{\x}{\x + 1} \right) \frac{\de\m_{-2}}{2} + \frac{\x}{\x + 1} \de\m_{-1} + \rig \\
\eq\lef + \frac{(1 \mp 1)(\hga_{-1} - 1)}{2} \frac{\x^2}{\x + 1} + \left( \frac{\g_\ph + \g_{\ph^5} - 2}{2} - \frac{\hga_{-2} - \hga_{-1}}{2} \frac{\x}{\x + 1} + \rig\rig \\
\eq\lef\lef \pm \frac{\g_{\ph} + \g_{\ph^5} - \hga_{-2} - \hga_{-1} - 1}{2} \frac{\x}{\x + 1} \right) \x^2\log\x + \rig \\
\eq\lef - \left[ (\hga_{-2} + \hga_{-1} - 2) \mp \left( \frac{\x}{\x + 1} + (\hga_{-2} - \hga_{-1}) \right) \right] \frac{\x^2\log(\x + 1)}{2} \right) \ .
\end{aligned}
\end{equation}

\bibliographystyle{utphys}
\footnotesize
\bibliography{References}	

\providecommand{\href}[2]{#2}\begingroup\raggedright\begin{thebibliography}{10}

\bibitem{SoderbergRousu:2023ucv}
A.~S\"oderberg~Rousu, {\em {Defects, renormalization and conformal field
  theory}}.
\newblock PhD thesis, Uppsala U., 2023.

\bibitem{Billo:2016cpy}
M.~Bill\`o, V.~Gon\c{c}alves, E.~Lauria, and M.~Meineri, ``{Defects in
  conformal field theory},''
  \href{http://dx.doi.org/10.1007/JHEP04(2016)091}{{\em JHEP} {\bfseries 04}
  (2016) 091}, \href{http://arxiv.org/abs/1601.02883}{{\ttfamily
  arXiv:1601.02883 [hep-th]}}.

\bibitem{Lauria:2018klo}
E.~Lauria, M.~Meineri, and E.~Trevisani, ``{Spinning operators and defects in
  conformal field theory},''
  \href{http://dx.doi.org/10.1007/JHEP08(2019)066}{{\em JHEP} {\bfseries 08}
  (2019) 066}, \href{http://arxiv.org/abs/1807.02522}{{\ttfamily
  arXiv:1807.02522 [hep-th]}}.

\bibitem{domb2000phase}
C.~Domb, {\em Phase transitions and critical phenomena}.
\newblock Elsevier, 2000.

\bibitem{PhysRevB.47.5841}
H.~W. Diehl and M.~Smock, ``Critical behavior at supercritical surface
  enhancement: Temperature singularity of surface magnetization and
  order-parameter profile to one-loop order,''
  \href{http://dx.doi.org/10.1103/PhysRevB.47.5841}{{\em Phys. Rev. B}
  {\bfseries 47} (Mar, 1993) 5841--5848}.

\bibitem{Shpot:2019iwk}
M.~A. Shpot, ``{Boundary conformal field theory at the extraordinary
  transition: The layer susceptibility to $O(\varepsilon)$},''
  \href{http://dx.doi.org/10.1007/JHEP01(2021)055}{{\em JHEP} {\bfseries 01}
  (2021) 055}, \href{http://arxiv.org/abs/1912.03021}{{\ttfamily
  arXiv:1912.03021 [hep-th]}}.

\bibitem{Liendo:2012hy}
P.~Liendo, L.~Rastelli, and B.~C. van Rees, ``{The Bootstrap Program for
  Boundary CFT$_d$},'' \href{http://dx.doi.org/10.1007/JHEP07(2013)113}{{\em
  JHEP} {\bfseries 07} (2013) 113},
  \href{http://arxiv.org/abs/1210.4258}{{\ttfamily arXiv:1210.4258 [hep-th]}}.

\bibitem{BELAVIN1984333}
A.~Belavin, A.~Polyakov, and A.~Zamolodchikov, ``Infinite conformal symmetry in
  two-dimensional quantum field theory,''
  \href{http://dx.doi.org/https://doi.org/10.1016/0550-3213(84)90052-X}{{\em
  Nuclear Physics B} {\bfseries 241} no.~2, (1984) 333--380}.

\bibitem{Rattazzi:2008pe}
R.~Rattazzi, V.~S. Rychkov, E.~Tonni, and A.~Vichi, ``{Bounding scalar operator
  dimensions in 4D CFT},''
  \href{http://dx.doi.org/10.1088/1126-6708/2008/12/031}{{\em JHEP} {\bfseries
  12} (2008) 031}, \href{http://arxiv.org/abs/0807.0004}{{\ttfamily
  arXiv:0807.0004 [hep-th]}}.

\bibitem{McAvity:1995zd}
D.~M. McAvity and H.~Osborn, ``{Conformal field theories near a boundary in
  general dimensions},''
  \href{http://dx.doi.org/10.1016/0550-3213(95)00476-9}{{\em Nucl. Phys. B}
  {\bfseries 455} (1995) 522--576},
  \href{http://arxiv.org/abs/cond-mat/9505127}{{\ttfamily
  arXiv:cond-mat/9505127}}.

\bibitem{Gliozzi:2015qsa}
F.~Gliozzi, P.~Liendo, M.~Meineri, and A.~Rago, ``{Boundary and Interface CFTs
  from the Conformal Bootstrap},''
  \href{http://dx.doi.org/10.1007/JHEP05(2015)036}{{\em JHEP} {\bfseries 05}
  (2015) 036}, \href{http://arxiv.org/abs/1502.07217}{{\ttfamily
  arXiv:1502.07217 [hep-th]}}.

\bibitem{Padayasi:2021sik}
J.~Padayasi, A.~Krishnan, M.~A. Metlitski, I.~A. Gruzberg, and M.~Meineri,
  ``{The extraordinary boundary transition in the 3d O(N) model via conformal
  bootstrap},'' \href{http://dx.doi.org/10.21468/SciPostPhys.12.6.190}{{\em
  SciPost Phys.} {\bfseries 12} no.~6, (2022) 190},
  \href{http://arxiv.org/abs/2111.03071}{{\ttfamily arXiv:2111.03071
  [cond-mat.stat-mech]}}.

\bibitem{Behan:2020nsf}
C.~Behan, L.~Di~Pietro, E.~Lauria, and B.~C. Van~Rees, ``{Bootstrapping
  boundary-localized interactions},''
  \href{http://dx.doi.org/10.1007/JHEP12(2020)182}{{\em JHEP} {\bfseries 12}
  (2020) 182}, \href{http://arxiv.org/abs/2009.03336}{{\ttfamily
  arXiv:2009.03336 [hep-th]}}.

\bibitem{Behan:2021tcn}
C.~Behan, L.~Di~Pietro, E.~Lauria, and B.~C. van Rees, ``{Bootstrapping
  boundary-localized interactions II. Minimal models at the boundary},''
  \href{http://dx.doi.org/10.1007/JHEP03(2022)146}{{\em JHEP} {\bfseries 03}
  (2022) 146}, \href{http://arxiv.org/abs/2111.04747}{{\ttfamily
  arXiv:2111.04747 [hep-th]}}.

\bibitem{Caron-Huot:2017vep}
S.~Caron-Huot, ``{Analyticity in Spin in Conformal Theories},''
  \href{http://dx.doi.org/10.1007/JHEP09(2017)078}{{\em JHEP} {\bfseries 09}
  (2017) 078}, \href{http://arxiv.org/abs/1703.00278}{{\ttfamily
  arXiv:1703.00278 [hep-th]}}.

\bibitem{Simmons-Duffin:2017nub}
D.~Simmons-Duffin, D.~Stanford, and E.~Witten, ``{A spacetime derivation of the
  Lorentzian OPE inversion formula},''
  \href{http://dx.doi.org/10.1007/JHEP07(2018)085}{{\em JHEP} {\bfseries 07}
  (2018) 085}, \href{http://arxiv.org/abs/1711.03816}{{\ttfamily
  arXiv:1711.03816 [hep-th]}}.

\bibitem{Bissi:2019kkx}
A.~Bissi, P.~Dey, and T.~Hansen, ``{Dispersion Relation for CFT Four-Point
  Functions},'' \href{http://dx.doi.org/10.1007/JHEP04(2020)092}{{\em JHEP}
  {\bfseries 04} (2020) 092}, \href{http://arxiv.org/abs/1910.04661}{{\ttfamily
  arXiv:1910.04661 [hep-th]}}.

\bibitem{Carmi:2019cub}
D.~Carmi and S.~Caron-Huot, ``{A Conformal Dispersion Relation: Correlations
  from Absorption},'' \href{http://dx.doi.org/10.1007/JHEP09(2020)009}{{\em
  JHEP} {\bfseries 09} (2020) 009},
  \href{http://arxiv.org/abs/1910.12123}{{\ttfamily arXiv:1910.12123
  [hep-th]}}.

\bibitem{Lemos:2017vnx}
M.~Lemos, P.~Liendo, M.~Meineri, and S.~Sarkar, ``{Universality at large
  transverse spin in defect CFT},''
  \href{http://dx.doi.org/10.1007/JHEP09(2018)091}{{\em JHEP} {\bfseries 09}
  (2018) 091}, \href{http://arxiv.org/abs/1712.08185}{{\ttfamily
  arXiv:1712.08185 [hep-th]}}.

\bibitem{Liendo:2019jpu}
P.~Liendo, Y.~Linke, and V.~Schomerus, ``{A Lorentzian inversion formula for
  defect CFT},'' \href{http://dx.doi.org/10.1007/JHEP08(2020)163}{{\em JHEP}
  {\bfseries 08} (2020) 163}, \href{http://arxiv.org/abs/1903.05222}{{\ttfamily
  arXiv:1903.05222 [hep-th]}}.

\bibitem{Gimenez-Grau:2021wiv}
A.~Gimenez-Grau and P.~Liendo, ``{Bootstrapping Monodromy Defects in the
  Wess-Zumino Model},'' \href{http://arxiv.org/abs/2108.05107}{{\ttfamily
  arXiv:2108.05107 [hep-th]}}.

\bibitem{Barrat:2022psm}
J.~Barrat, A.~Gimenez-Grau, and P.~Liendo, ``{A dispersion relation for defect
  CFT},'' \href{http://arxiv.org/abs/2205.09765}{{\ttfamily arXiv:2205.09765
  [hep-th]}}.

\bibitem{Bianchi:2022ppi}
L.~Bianchi and D.~Bonomi, ``{Conformal dispersion relations for defects and
  boundaries},'' \href{http://arxiv.org/abs/2205.09775}{{\ttfamily
  arXiv:2205.09775 [hep-th]}}.

\bibitem{Kaviraj:2018tfd}
A.~Kaviraj and M.~F. Paulos, ``{The Functional Bootstrap for Boundary CFT},''
  \href{http://dx.doi.org/10.1007/JHEP04(2020)135}{{\em JHEP} {\bfseries 04}
  (2020) 135}, \href{http://arxiv.org/abs/1812.04034}{{\ttfamily
  arXiv:1812.04034 [hep-th]}}.

\bibitem{Mazac:2018biw}
D.~Maz\'a\v{c}, L.~Rastelli, and X.~Zhou, ``{An analytic approach to
  BCFT$_{d}$},'' \href{http://dx.doi.org/10.1007/JHEP12(2019)004}{{\em JHEP}
  {\bfseries 12} (2019) 004}, \href{http://arxiv.org/abs/1812.09314}{{\ttfamily
  arXiv:1812.09314 [hep-th]}}.

\bibitem{Bissi:2018mcq}
A.~Bissi, T.~Hansen, and A.~S\"oderberg, ``{Analytic Bootstrap for Boundary
  CFT},'' \href{http://dx.doi.org/10.1007/JHEP01(2019)010}{{\em JHEP}
  {\bfseries 01} (2019) 010}, \href{http://arxiv.org/abs/1808.08155}{{\ttfamily
  arXiv:1808.08155 [hep-th]}}.

\bibitem{Dey:2020jlc}
P.~Dey and A.~S\"oderberg, ``{On analytic bootstrap for interface and boundary
  CFT},'' \href{http://dx.doi.org/10.1007/JHEP07(2021)013}{{\em JHEP}
  {\bfseries 07} (2021) 013}, \href{http://arxiv.org/abs/2012.11344}{{\ttfamily
  arXiv:2012.11344 [hep-th]}}.

\bibitem{Gimenez-Grau:2020jvf}
A.~Gimenez-Grau, P.~Liendo, and P.~van Vliet, ``{Superconformal boundaries in
  $4-\epsilon$ dimensions},''
  \href{http://dx.doi.org/10.1007/JHEP04(2021)167}{{\em JHEP} {\bfseries 04}
  (2021) 167}, \href{http://arxiv.org/abs/2012.00018}{{\ttfamily
  arXiv:2012.00018 [hep-th]}}.

\bibitem{Lauria:2017wav}
E.~Lauria, M.~Meineri, and E.~Trevisani, ``{Radial coordinates for defect
  CFTs},'' \href{http://dx.doi.org/10.1007/JHEP11(2018)148}{{\em JHEP}
  {\bfseries 11} (2018) 148}, \href{http://arxiv.org/abs/1712.07668}{{\ttfamily
  arXiv:1712.07668 [hep-th]}}.

\bibitem{Alday:2021ajh}
L.~F. Alday, A.~Bissi, and X.~Zhou, ``{One-loop gluon amplitudes in AdS},''
  \href{http://dx.doi.org/10.1007/JHEP02(2022)105}{{\em JHEP} {\bfseries 02}
  (2022) 105}, \href{http://arxiv.org/abs/2110.09861}{{\ttfamily
  arXiv:2110.09861 [hep-th]}}.

\bibitem{Huber:2005yg}
T.~Huber and D.~Maitre, ``{HypExp: A Mathematica package for expanding
  hypergeometric functions around integer-valued parameters},''
  \href{http://dx.doi.org/10.1016/j.cpc.2006.01.007}{{\em Comput. Phys.
  Commun.} {\bfseries 175} (2006) 122--144},
  \href{http://arxiv.org/abs/hep-ph/0507094}{{\ttfamily arXiv:hep-ph/0507094}}.

\bibitem{Huber:2007dx}
T.~Huber and D.~Maitre, ``{HypExp 2, Expanding Hypergeometric Functions about
  Half-Integer Parameters},''
  \href{http://dx.doi.org/10.1016/j.cpc.2007.12.008}{{\em Comput. Phys.
  Commun.} {\bfseries 178} (2008) 755--776},
  \href{http://arxiv.org/abs/0708.2443}{{\ttfamily arXiv:0708.2443 [hep-ph]}}.

\bibitem{Rychkov:2015naa}
S.~Rychkov and Z.~M. Tan, ``{The $\epsilon$-expansion from conformal field
  theory},'' \href{http://dx.doi.org/10.1088/1751-8113/48/29/29FT01}{{\em J.
  Phys. A} {\bfseries 48} no.~29, (2015) 29FT01},
  \href{http://arxiv.org/abs/1505.00963}{{\ttfamily arXiv:1505.00963
  [hep-th]}}.

\bibitem{Herzog:2017xha}
C.~P. Herzog and K.-W. Huang, ``{Boundary Conformal Field Theory and a Boundary
  Central Charge},'' \href{http://dx.doi.org/10.1007/JHEP10(2017)189}{{\em
  JHEP} {\bfseries 10} (2017) 189},
  \href{http://arxiv.org/abs/1707.06224}{{\ttfamily arXiv:1707.06224
  [hep-th]}}.

\bibitem{Herzog:2022jlx}
C.~P. Herzog and V.~Schaub, ``{Fermions in Boundary Conformal Field Theory :
  Crossing Symmetry and $\epsilon$-Expansion},''
  \href{http://arxiv.org/abs/2209.05511}{{\ttfamily arXiv:2209.05511
  [hep-th]}}.

\end{thebibliography}\endgroup
	
\end{document}